# Manipulating magnetization by orbital current from a light metal Ti

Dongxing Zheng[1], Jingkai Xu[1], Fatimah Alsayafi[1], Sachin Krishnia[2], Dongwook Go[2,3], Duc Tran[2], Tao Yang[1], Yan Li[1], Yinchang Ma[1], Chen Liu[1], Meng Tang[1], Aitian Chen[1], Hanin Algaidi[1], Hao Wu[4], Kai Liu[5], Yuriy Mokrousov[2,3], Mathias Kläui[2], Udo Schwingenschlögl[1], and Xixiang Zhang[1*]

[1]Physical Science and Engineering Division, King Abdullah University of Science and Technology (KAUST), Thuwal 23955–6900, Saudi Arabia

[2]Institute of Physics, Johannes Gutenberg University Mainz, 55099, Mainz, Germany

[3]Peter Grünberg Institut, Forschungszentrum Jülich, 52425 Jülich, Germany

[4]Songshan Lake Materials Laboratory, Dongguan, Guangdong 523808, China

[5]Physics Department, Georgetown University, Washington, DC 20057, USA

Dongxing Zheng, Jingkai Xu and Fatimah Alsayafi contributed equally to this work.

[*]Corresponding authors:

Xixiang Zhang (email: xixiang.zhang@kaust.edu.sa)

**Abstract**

The orbital Hall effect, which does not rely on the spin-orbit coupling, has recently emerged as a promising mechanism for electrically manipulating magnetization in thin-film ferromagnets. Despite its potential, direct experimental observation of magnetization switching driven by orbital currents has been challenging, primarily because there is no direct exchange coupling between orbital angular momentum and local spin based magnetic moments. In this study, we present a compensated design to directly probe the contribution of orbital currents in the most promising light metal titanium (Ti), where symmetric layer structures allow zeroing out of the net spin current. By varying the thickness of the Ti layer in Ti($t$)/Pt/Co/Pt/Co/Pt multilayers, we demonstrate the ability to control the magnetization switching polarity. We deduce the orbital charge conversion efficiency of the Ti layer to be approximately 0.17. These findings not only confirm the presence of the orbital Hall effect in Ti but also suggest that orbital currents may be promising candidates for developing energy-efficient magnetic devices with enhanced performance and scalability.

Spin angular momentum (*S*) and orbital angular momentum (*L*) are intrinsic properties of electrons within atoms. While spin angular momentum can be generated through mechanisms such as the Spin Hall effect (SHE) or the Rashba-Edelstein effect,[1-9] orbital angular momentum has often been overlooked, primarily due to orbital quenching in equilibrium by the crystal field or electron localization.[10-13] Recent studies, however, have demonstrated that an orbital texture can produce finite orbital angular momentum even in the absence of spin-orbit coupling (SOC) when subjected to an electric field.[14-26] Despite this progress, one remaining issue is that there is no direct exchange coupling between the orbital angular momentum and the local spin-based magnetic moments, and thus the orbital torque cannot directly give a torque on magnetization;[12,13,25-30] Therefore, the SOC is required as a mediator, which can convert the orbital angular momentum into spin angular momentum by introducing a nonmagnetic metallic layer with strong SOC or generate a spin-orbit effective field by utilizing the SOC within the ferromagnetic layer itself (see Fig. S2 in Supplemental Material).[29-35] The necessity of SOC for this conversion complicates the distinction between orbital torque induced by the orbital Hall effect (OHE) and spin torque induced by the SHE. Consequently, although the orbital torque has been widely studied, the direct switching of magnetization solely by orbital torque remains a subject of ongoing debate.

In this study, we propose a compensated design to isolate the contribution of orbital angular momentum, as illustrated in Fig. 1a. The ferromagnetic layer is sandwiched between two Pt layers with strong SOC, enabling control of the effective spin current generated by these Pt layers and its impact on the ferromagnetic layer by adjusting the thickness of the top and bottom Pt layers. This approach allows us to achieve states where the spin current is dominated by the bottom layer, or the top layer, or stays neutralized. Furthermore, as shown in Fig. 1b, we further refine these compensated states by inserting Ti layer. By varying the Ti layer thickness, we effectively isolate and identify the contribution of orbital torque from the light metal Ti, demonstrating the orbital origin of the strong torques. The orbital current generated in the Ti layer can be converted into spin current by the strong SOC of the Pt layer, which subsequently interacts with the local spin-based moment in the FM layer and transfers its spin angular momentum to the FM layer directly, as is schematically shown in Fig. 1(c).

Based on this design, we have prepared two series of multilayers: Ti/Pt/Co/Pt/Co/Pt with perpendicular magnetic anisotropy and Ti/Pt/Py/Pt with in-plane magnetic anisotropy, using a Singulus ROTARIS magnetron sputtering system. We have successfully demonstrated current-induced switching of perpendicular magnetization driven by the orbital current generated in the light metal Ti within the Ti/Pt/Co/Pt/Co/Pt multilayers. In these Ti/Pt/Co/Pt/Co/Pt multilayers, we observe that the current-induced magnetization switching polarity could be reversed from clockwise to anticlockwise, signifying a change in the injected spin polarization within the ferromagnetic layer. This reversal of spin polarization was validated through second harmonic voltage measurements in the Ti/Pt/Py/Pt multilayers. Our findings reveal the impact of orbital currents to the torques and their pivotal role in the emerging field of orbitronics.

To investigate the compensated states of the spin current generated by the Pt(2.0)/Co(0.4)/Pt(0.6)/Co(0.4)/Pt($t_{top}$) (PtCoPt) multilayers, we fabricated a series of such multilayers with varying the top Pt layer thicknesses (in nanometers). The structure of the PtCoPt($t_{top}$) multilayers is schematically illustrated in the inset of Fig. 2a. Additional fabrication details are provided in the Methods section of the Supplemental Material. The anomalous Hall effect (AHE) measured by a perpendicular magnetic field, shown in Fig. 2a, reveal square-shaped AHE loops, confirming the perpendicular magnetic anisotropy of the PtCoPt($t_{top}$) multilayers.

We then conduct current-induced magnetization switching measurements on these multilayers. As shown in Fig. 2b, the switching polarity of the PtCoPt($t_{top}$) multilayers is found to depend on the thickness of the top Pt layer, $t_{top}$. When $t_{top}$=1.2 nm, the switching polarity at $H_x$=360 Oe is clockwise, indicating that the net spin current was dominated by the bottom Pt layer, consistent with the positive spin Hall angle of Pt.[3,5,36] In contrast, when the $t_{top}$ was increased to 3.0 nm, the switching polarity reversed, suggesting that the net spin current was now dominated by the top Pt layer. The reversal of the switching polarity was further evidenced by the $t_{top}$ dependent switching ratio. As shown in Fig. 2c, at an external field $H_x$=360 Oe, the switching ratio gradually shifts from positive to negative as $t_{top}$ increases, while it shifts from negative to positive with $H_x$=-360 Oe. The crossover around $t_{top}$=1.9 nm indicates a fully compensated state of the PtCoPt($t_{top}$) multilayers. Note that the compensation is not at exactly identical thicknesses for the top and bottom Pt layers. This might be due to the different substrates on which the Pt layer was grown on. The opposite switching ratios observed at $H_x$=360 and -360 Oe are attributed to the opposite sign of the external torque generated by $H_x$.[4,5] Additionally, the $t_{top}$-dependent critical switching current, which peaks around 1.9 nm as shown in the inset of Fig. 2c, corroborates the nearly compensated state of the spin current at this thickness.

After confirming the controllable net spin current by varying the thickness of the top Pt layer, we introduced a 3.0 nm thick Ti layer beneath the bottom Pt layer to investigate the influence of orbital current generated by the Ti layer on magnetization switching. The Ti(3.0)/PtCoPt($t_{top}$) multilayers maintains the perpendicular magnetic anisotropy after the insertion of a 3.0 nm thick Ti layer, as confirmed by the square shape like AHE loop shown in Fig. 2d. We then perform current-induced magnetization switching measurements and observe the following key features: Firstly, the thickness-dependent switching behavior of the top Pt layer was significantly altered by the insertion of the Ti layer. As shown in Fig. 2e, the switching polarity for Ti(3.0)/PtCoPt(1.2) is anticlockwise, which is opposite to the clockwise switching polarity observed in PtCoPt(1.2) multilayers (Fig. 2b). This result demonstrates that the Ti layer plays a crucial role in compensating the spin current generated by the top and bottom Pt layers. Secondly, the additional contribution of the Ti layer is further confirmed by the thickness-dependent switching ratio shown in the Fig. 2f. The crossover of the switching ratio as a function of $t_{top}$ under external field $H_x$ of 360 Oe and -360 Oe occurs around 1.0 nm, which is significantly lower than the value of ~1.9 nm in the PtCoPt($t_{top}$) multilayers. Furthermore, the critical switching current was also found to be influenced by the presence of the Ti layer. As is shown in the inset of Fig. 2e, the critical switching current initially increases with the increasing $t_{top}$, reaching a maximum around 1.0 nm. As $t_{top}$ increases further, the critical switching current decreases, reaching a minimum around 1.3 nm, and then rises again with the increasing $t_{top}$. The initial increase is due to the decrease in net spin current as $t_{top}$ increases, which compensates the total spin current generated by the bottom Pt layer and the orbital current generated by the Ti layer. When the top Pt layer thickness reaches around 1 nm, the spin current is nearly compensated, rendering current-induced magnetization switching ineffective. However, as $t_{top}$ continues to increase, the net spin current increases again, leading to a decrease in the critical switching current. After reaching a minimum around $t_{top}$=1.3 nm, further increasing the Pt layer thickness does not favor a further reduction in the critical switching current due to the current shunting effect,[37,38] which therefore requires a higher switching current to maintain the same switching current density.

The results above indicate that the Ti layer inserted beneath the bottom Pt layer has a similar contribution to the net spin current as the top Pt layer. This finding is further confirmed by studying the Ti layer thickness dependence in the Ti($t$)/PtCoPt multilayers, with the bottom and top Pt layer thickness fixed at 2.0 and 1.5 nm, respectively. Figure 3a presents a mapping of the Hall resistance $R_H$ as a function of applied current and external magnetic field $H_x$ of the Ti(2.5)/PtCoPt multilayer. In this mapping, red, purple, and blue colors represent up, partial, and down magnetization states, respectively. The mapping results demonstrate that the magnetization can be efficiently switched by the torques generated by the Pt and Ti layers, with switching polarity correlated with the applied magnetic field.

Figures 3b-d show the current-induced magnetization switching in the Ti($t$)/PtCoPt multilayers with Ti layer thickness of 0.5, 1.0 and 1.5 nm under an in-plane magnetic field $H_x$=360 Oe. The switching polarity is found to depend on the thickness of the Ti layer. In Fig. 3b, when the inserted Ti layer is thin of 0.5 nm, the switching polarity of Ti(0.5)/PtCoPt(1.5) multilayer is found to be the same as that of the PtCoPt(1.2) multilayer (Fig. 2b), where the net spin current is dominated by the bottom Pt layer. This suggests that the net spin current is still dominated by the 2 nm thick bottom Pt layer. When the thickness of the Ti layer is increased to 1.0 nm (Fig. 3c), the magnetization switching is nearly impossible to see, indicating that the net spin current in the PtCoPt(1.5) multilayer is nearly compensated by the Ti layer. As the Ti layer thickness increases further to 1.5 nm (Fig. 3d), the switching polarity reverses to anticlockwise, signifying a change in the net spin current's spin polarization direction from +$y$ to -$y$. Given that the PtCoPt(1.5) multilayer remains unchanged, this reversal is attributed to the change in Ti layer thickness, which is reported to have a strong orbital Hall effect and can influence the magnetization switching.[24,33,39,40]

This behavior is also reflected in the thickness-dependent switching ratio and critical switching current. As shown in the inset of Fig. 3e, the critical switching current initially increases with Ti layer thickness, which corresponds to a decrease in the net spin current. When the Ti layer reaches 1.0 nm, magnetization switching becomes nearly undetectable, marking the compensation point. Beyond this point, the critical switching current decreases with further Ti thickness increases, corresponding to an increase in net spin current. The switching ratio exhibits a similar trend. At an external field of $H_x$=360 Oe, the switching ratio gradually shifts from positive to negative, while it shifts from negative to positive with $H_x$=-360 Oe. The cross over around $t$=1.0 nm indicates a fully compensated state in the Ti(1.0)/PtCoPt(1.5) multilayers. Once the Ti thickness surpasses the compensation point, the switching ratio increases gradually.

Having observed the sign change of the switching, we next need to confirm that the underlying torques are the origin of this sign change. So to confirm the contribution of orbital current from the light metal Ti, we performed second harmonic Hall voltage measurement on the Ti($t$)/Pt(2)/Py(4)/Pt(1) series multilayers. The second Harmonic voltage $V_{2\omega}$ as a function of the angle $\varphi$ is shown in Fig. 4a, which is obtained by rotating the sample in the $xy$ plane with a fixed external magnetic field. The details about the measurements and analysis can be found in the Supplemental Material. Figure 4a shows the angular dependence of second harmonic Hall voltage $V_{2\omega}$ for the Pt(2)/Py(4)/Pt(1) and Pt(2)/Py(4)/Pt(3) multilayers, where the net spin current generated by the applied AC current is primarily injected into the Py layer by the bottom and top Pt layer, respectively. When a spin current is injected into a ferromagnet layer, two types of torques will affect the dynamics of magnetization in the Py layer: a field-like torque $\sim \vec{m} \times \vec{\sigma}$ and a damping-like torque $\sim \vec{m} \times (\vec{\sigma} \times \vec{m})$.[7,41-43] Here, $\vec{\sigma}$ is the accumulated spin direction and $\vec{m}$ is the normalized magnetization vector. The magnetization of the ferromagnetic layer oscillates at the same frequency as the applied

AC current around its equilibrium position, resulting in a Hall voltage that contains a second harmonic component directly related to the damping-like and field-like torques.[44-47] As we discussed above, in the Pt(2)/Py(4)/Pt(1) and Pt(2)/Py(4)/Pt(3) multilayers, the net spin current should be dominated by the bottom and top Pt layers respectively. Therefore, spin currents with opposite spin polarization are expected to be injected into the Py layer. Consequently, the second harmonic Hall voltage generated in these two multilayers should have opposite signs.[44,45] As shown in Fig. 4a, our results confirm this expectation, with the second harmonic Hall voltage exhibiting opposite signs in the Pt(2)/Py(4)/Pt(1) and Pt(2)/Py(4)/Pt(3) multilayers. This finding demonstrates that the second harmonic voltage measurement is an effective method for characterizing the sign change of the injected spin current in the ferromagnetic layer.

We then performed the second harmonic Hall voltage measurement for the Ti($t$)/Pt(2)/Py(4)/Pt(1) multilayers. Figure 4b shows the second harmonic Hall voltage for the Ti($t$)/Pt(2)/Py(4)/Pt(1) multilayers with the Ti layer thickness of 0.5 and 2.5 nm. In the Ti(0.5)/Pt(2)/Py(4)/Pt(1) multilayer, the angular dependence of the second harmonic Hall voltage is similar to that of the Pt(2)/Py(4)/Pt(1) multilayer, with the same sign of the second harmonic Hall voltage. However, when the Ti layer thickness is increased to 2.5 nm, the sign of the second harmonic Hall voltage reverses, consistent with the reversed switching polarity observed in the Ti($t$)/PtCoPt multilayers. The second harmonic Hall voltage correlated to the damping like torque term is shown in Fig. 4c. The slopes of linear fits to the $V_{DL}$ as a function of $1/(H_k\text{-}H_{ext})$ give the information about effective field $H_{DL}$ that is attributed to the damping like torque, $H_k$ is the out-of-plane effective magnetic field.[44,45,48] We estimate the charge-to-spin conversion efficiency by using the formula $\xi_{\text{DL}} = \frac{2e\mu_0 M_s t_{\text{FM}} H_{\text{DL}}}{\hbar J_{\text{ac}}}$,[4] where $e$ is the electron charge, $\hbar$ is the reduced Plank constant, $M_s$ and $t_{\text{FM}}$ is the saturation magnetization and thickness of the Py layer, $J_{\text{ac}}$ is the applied current density. The calculated effective spin Hall angle and effective orbital Hall angle values are ~0.05 for the Pt and ~0.17 for the Ti in the Ti($t$)/Pt(2)/Py(4)/Pt(1) multilayers, which shows the strong orbital Hall effect consistent with previous reports.[33,39,49-54]

In conclusion, we conducted a systematic study on magnetization switching induced by orbital current in the Ti($t$)/PtCoPt multilayers and on orbital torque characterization in the Ti($t$)/Pt/Py/Pt multilayers. By developing a compensated design, we were able to counteract the net contribution of spin current from the spin Hall effect of the Pt layer by varying the thickness of the Ti layer, which leads to a reversal of the switching polarity in the Ti($t$)/PtCoPt multilayers. This observation was further validated by sign reversal of the second harmonic Hall voltage in the Ti($t$)/Pt/Py/Pt multilayers. The orbital Hall angle of the Ti layer in these multilayers is quantified to be 0.17. Our findings not only successfully isolate and identify the contribution of orbital currents but also provide a novel pathway for the development of spin-orbit torque-based memory technologies making use of the very strong orbital Hall effect found in our stacks.

**References and Notes**

**Acknowledgments:**

This work is supported by the King Abdullah University of Science and Technology, Office of Sponsored Research (OSR), under award Nos. ORA-

CRG10-2021-4665; ORA-CRG11-2022-5031 and ORA-CRG8-2020-4048. The study also has been supported by the European Horizon Europe Framework Programme under an EC Grant Agreement N101129641 "OBELIX". We also gratefully acknowledge financial support by the Deutsche Forschungsgemeinschaft (DFG, German Research Foundation) – TRR 173/2 – 268565370 (projects A11 and A01).

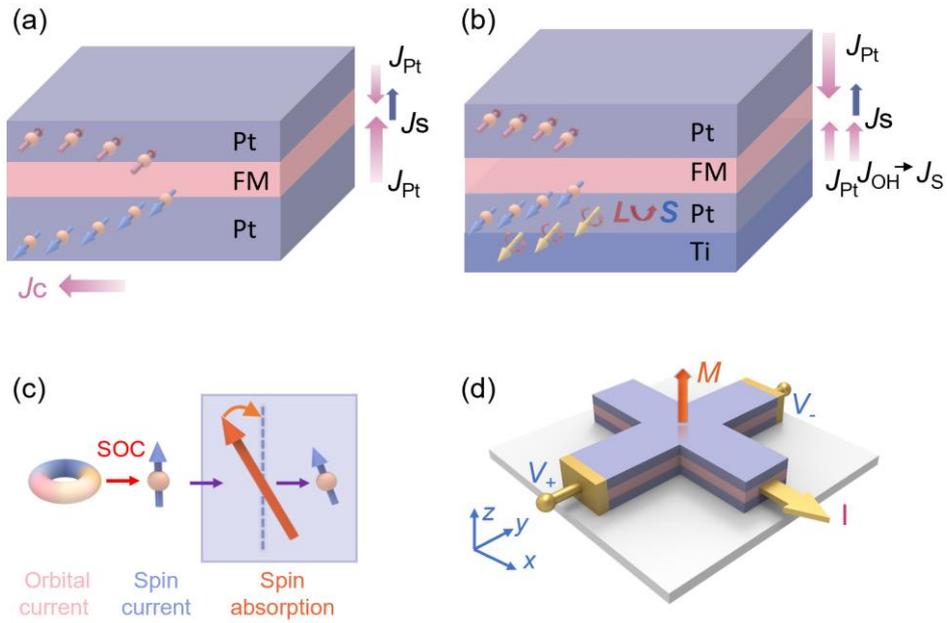

**Fig. 1. Schematic of the compensated design and measurement setup.** (**a**) A ferromagnetic (FM) layer is sandwiched between two Pt layers with varying thickness, allowing for the generation of a net spin current that acts on the FM layer. The magnitude and direction of the net spin current can be adjusSted by controlling the relative thicknesses of the top and bottom Pt layers. (**b**) The net spin current in the Ti/Pt/FM/Pt multilayer can be either compensated or enhanced by the orbital current generated in a light metal Ti layer with weak spin-orbit coupling, providing additional control over the system's magnetic behavior. (**c**) Schematic illustration of the orbital-to-spin current conversion mediated by the strong spin-orbit coupling. The converted spin current exerts a torque on the magnetization of the FM layer. (**d**) Schematic of the experimental setup for current-induced magnetization switching measurements.

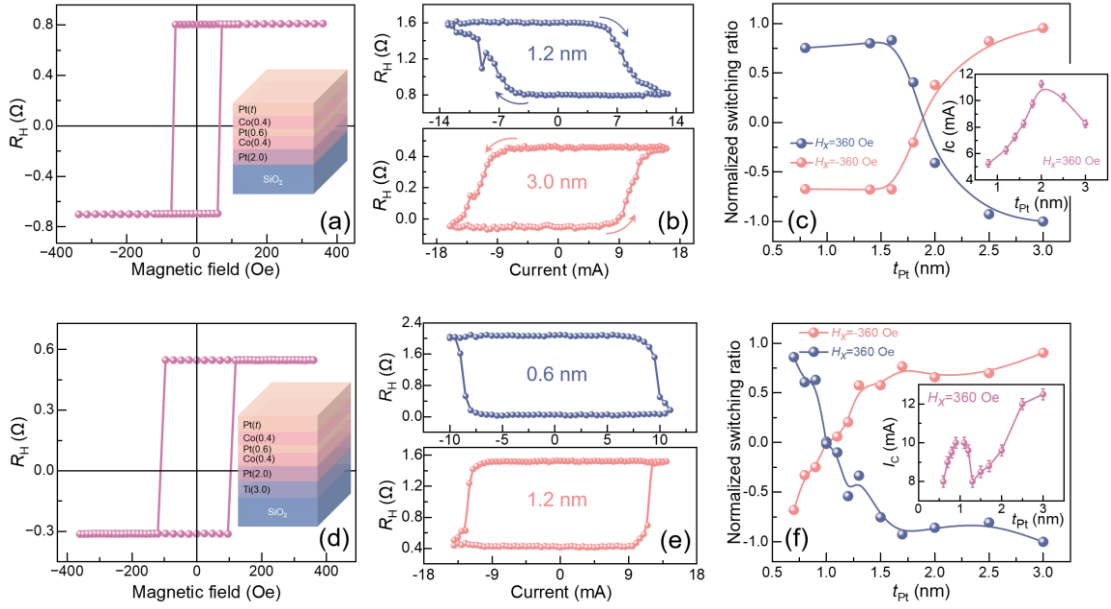

**Fig. 2. Orbital current induced magnetization switching.** AHE loops for the PtCoPt($t_{top}$) multilayer (**a**) and Ti(3.0)/PtCoPt($t_{top}$) multilayer (**d**), the thickness of bottom Pt layer is fixed at 2 nm. The insets schematically illustrate the structures of the multilayers. Current-induced magnetization switching in the PtCoPt($t_{top}$) multilayers with $t_{top}$ of 1.2 and 3.0 nm (**b**), and in the Ti(3.0)/PtCoPt($t_{top}$) multilayers with top $t_{top}$ of 0.6 and 3.0 nm (**e**), where a magnetic field $H_x$=360 Oe was applied along the current direction. Normalized switching ratio as a function of $t_{top}$ in the PtCoPt($t_{top}$) multilayers (**c**) and in the Ti(3.0)/PtCoPt($t_{top}$) multilayers (**f**). The current-induced magnetization switching ratio is calculated by determining the percentage ratio of |Δ$R_H$/ to $R_{AHE}$, with the positive and negative signs representing the clockwise and anticlockwise magnetization switching polarities. |Δ$R_H$/ is the amplitude of Hall resistance that is switched by the current, $R_{AHE}$ is the amplitude of the anomalous Hall resistance. The insets show the corresponding critical switching current as a function of $t_{top}$.

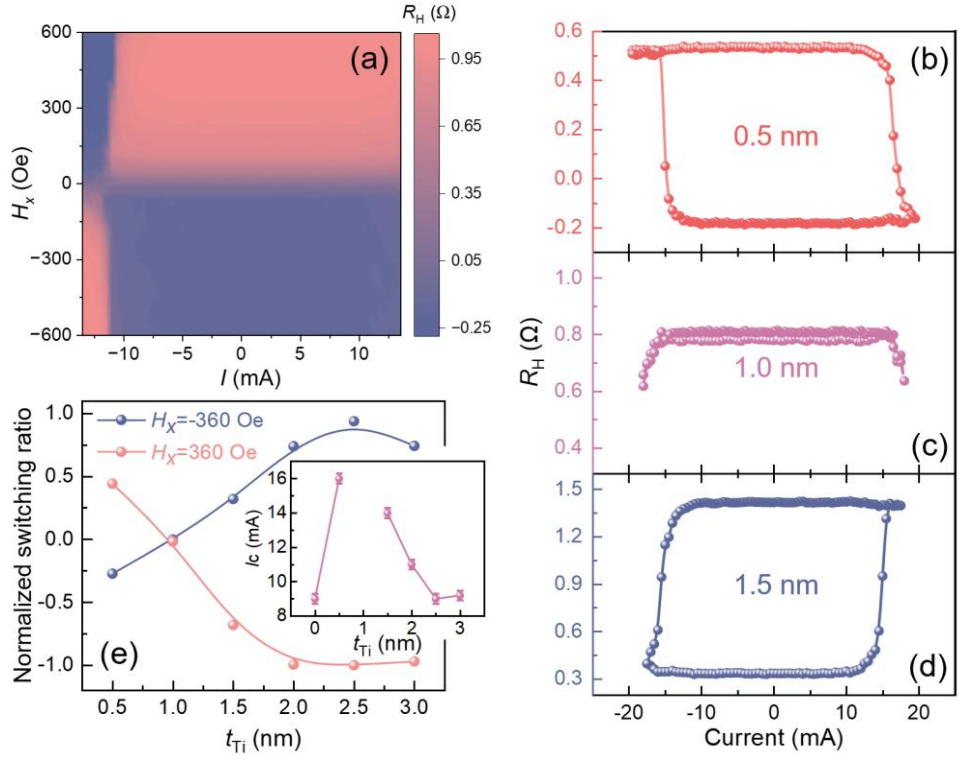

**Fig. 3. Ti thickness dependent magnetization switching.** (**a**) Phase diagram of the Hall resistance $R_H$ as a function of the applied current and magnetic field in the Ti(2.5)/PtCoPt(1.5) multilayer. Current-induced magnetization switching in the Ti($t$)/PtCoPt(1.5) multilayers with Ti layer thicknesses of 0.5 nm (**b**), 1.0 nm (**c**) and 1.5 nm (**d**). A magnetic field $H_x$=360 Oe was applied along the current direction. (**e**) Normalized switching ratio as a function of the thickness of the Ti layer in the Ti($t$)/PtCoPt(1.5) multilayers. The insets show the corresponding critical switching current as a function of the Ti layer thickness.

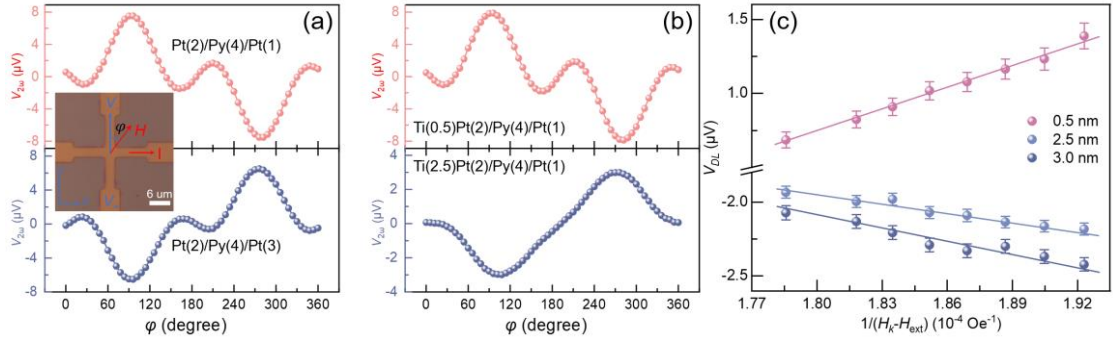

**Fig. 4. Spin orbit torque characterization.** (**a**) Second-harmonic Hall voltage $V_{2\omega}$ as a function of in-plane external magnetic field at 300 Oe for the Pt(2)/Py(4)/Pt(1) and Pt(2)/Py(4)/Pt(3) multilayers. The inset is the microscope image of the Hall bar device. (**b**) $V_{2\omega}$ for the Ti($t$)/Pt(2)/Py(4)/Pt(1) multilayers with Ti thickness of 0.5 and 2.5 nm. (**c**) Extracted damping-like torque contribution $V_{DL}$ as a function of the inverse of the anisotropy field, subtracting the external magnetic field.